**1. Title of the Project:**
**2. Theme(mention the domains as mentioned in the advertisement)**
**3. Project Category:** Category A
**3. b Proposed Duration of the Project:** Months

**4a. Principal Investigator (PI)**

| | |
|---|---|
| Name | **Dr. Mukesh Rawat** |
| Designation | **Professor** |
| Department | **Computer Science and engineering** |
| Institution | **Meerut Institute of engineering and technology** |
| Postal Address | |
| E-mail | **Mukesh.rawat@miet.ac.in** |
| Date of Birth: | |

**4b. Co-Principal Investigator (Co-PI)**

| | |
|---|---|
| Name | **Dr. Neha Kaushik** |
| Designation | **Lecturer** |
| Department | **Computer Science and engineering** |
| Institution | Kasturba Institute of Technology |
| Postal Address | |
| E-mail | neha.kaushik45@gov.in |
| Date of Birth: | |

**6. Context/Background** (250 words)

Indian Railways has got a strong, efficient system for grievance redressal. The system is named as Rail Madad[1]. Complaints redressal system works through the modes as given below:
1. Mobile App (Android/iOS)
2. Helpline Number (call as well as SMS)
3. Social Media(Facebook, twitter)
4. Email
5. Manual Dak

Rail Madad can be availed by any railway customer to raise complaints regarding Indian Railways service delivery. Rail Madad is very well automated and robust system for grievance redressal using Mobile App and Helpline Number. It is observed from the case study given in Footnote, that the grievances received on Social Media Accounts of Indian Railways are analyzed manually for redressal. It is observed that the system can further be improved by incorporating a plugin for automatic screening of grievances that the customers post on Social Media accounts of Indian Railways. The texts posted on the social media accounts can be processed using various text analysis techniques to identify the actionable tasks. We propose to build a plugin for identification of completeness of information posted by the customers and further to process the grievance into actionable tasks and thus reducing the human intervention and improving the efficiency of Rail Madad.

**7. Problems to be addressed**(1500 words)
The current grievance redressal system has a dedicated 24X7 Twitter Cell, wherein the human experts take actions and respond to the tweets of customers addressed to Ministry of Railways. It is done quite promptly by the human experts. It is understood that the software plugin to process the tweets addressed towards Ministry of Railways can not match the human expertise. Still, efforts can be done to build a software plugin which can ease the human effort.
Following types of tweets can be seen addressed to Ministry of Railways:
1. Complaints

---

[1] https://nceg.gov.in/sites/default/files/Rail%20Madad.pdf

Figure 1 shows an example complaint. In this tweet, one user has reported about water leakage at `bhandup` railway station `platform no 2/3`. This is just one complaint, where a negative word (we define negative word as a word which carries some type of malfunctioning information) 'leakage' can be identified. Further analysis of text reveals the location of the problem. But in many cases, the customers do not provide complete information needed to service the complaint. For example, in Figure 2, the customer has filed a complaint regarding 'ticket pnr not generated' and the customer wants a refund, which has not yet been done. To process the complaint, Railways need some information like *mobile number*, *user id*, *date of booking* and *transaction id*. We identify such tweets as incomplete information. It can be made out that the tweet is a complaint from the textual analysis and identification of the phrase 'ticket pnr not generated' and 'refund'. However, identifying that the tweet has incomplete information is quite a challenging task.

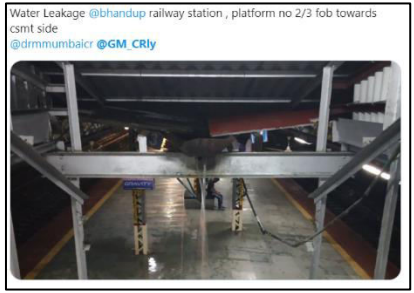
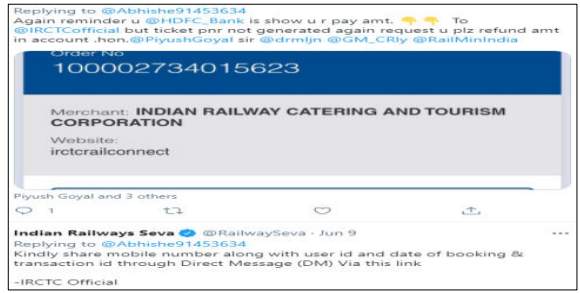

Figure 1 Example Complaint 1

Figure 2 Example Complaint 2

2. Suggestions

Second types of tweets are suggestions from the customers or twitter users addressed to Railways. Figure 3 shows an example suggestion tweet. In this tweet, the user/customer has suggested to include one more coach to the train so as to deal with the huge rush in the train. The user has also suggested to run some more specials from GKP(Gorakhpur). The good thing about this tweet is that the user has started the tweet with the word 'suggestion', which makes it clear even before reading the post that, what follows is a suggestion. But, it is not the case with all the users. Figure 4 shows such a scenario wherein identifying the tweet as suggestion is not this straightforward.

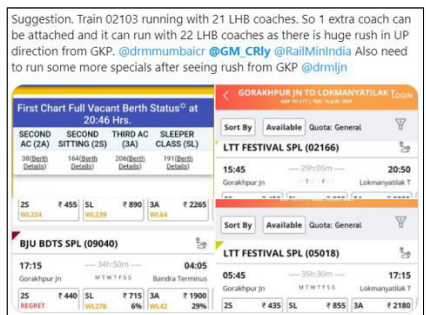
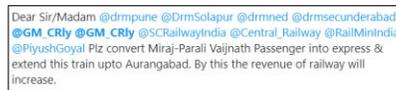

Figure 4 Example Suggestion 2

3. Appreciation

Figure 3 Example Suggestion1

Many a times, customers just share their appreciation for the Indian Railways through tweets. They either want to acknowledge the services provided, or the promptness in grievance redressal or any other thing that they may like during their journey. Figure 5 shows an example in which a customer has shown his/her appreciation for the scenic beauty the he/she is enjoying while travelling in Indian Railways. Figure 5 illustrates an example, where a customer has acknowledged the prompt response of Indian Railways towards his complaint.

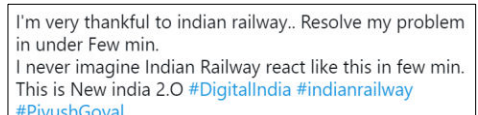

Figure 6 Example Appreciation 2

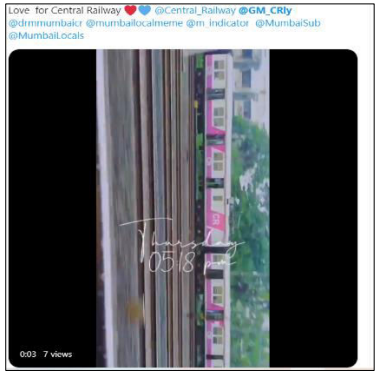
Figure 5 Example Appreciation 1

The problem areas identified during the preliminary analysis of the tweets data collected from twitter handles of Indian Railways and its officials are highlighted below:
a. Identification of the type of tweet
   This involves identifying whether the tweet is a complaint, suggestion or appreciation.
b. Identification of incomplete information
   If the tweet is a complaint, it is essential to identify whether the tweet contains all the required information for grievance redressal.
c. Identification of actionable tasks
   In this, we plan to identify the tasks that can be done to address the complaint(s) raised by the customer. This is quite challenging to automate this using textual analysis, as different customers use different language and semantics to write their complaints.
d. Identification of the division, region and department for Redressal
   Once the complaint has and actionable task has been identified, it can be directed to the relevant division, region or department based upon the existing structure of Railways grievance redressal system.

## 8. Aims and Objectives

This project aims at building a software plug-in to minimize the human effort involved in analysis of tweets addressed to Indian Railways and aid in existing complaints redressal system by identifying the complaints from the tweets. It is understood that it is not possible to match human promptness in terms of handling the tweets, still we can try to reduce the human efforts by working on the following objectives:
a. Identification of the Type of Tweets
   As outlined in the previous Section, tweets addressed to Indian Railways are categorised into three categories viz. complaints, suggestion and appreciation. The easiest way out for categorization of tweets into any one of the types is to instruct all the customers to prefix their tweets with a label, COMPLAINT for complaint, SUGGESTION for suggestion and APPRECIATION for appreciation tweets.
   However, it is understood that it is not possible to make this convention mandatory and moreover, Indian Railways is committed to provide quality service at customer's level of ease. Hence, we propose a keyword based approach for categorization of tweets into one of the type of tweets. In this approach, we label the words as positive, negative or neutral. If a tweet consists majorly of the negative and neutral words, it is considered as a complaint. If a tweet consists majorly of positive and neutral words, it is considered as an appreciation. If a tweet consists majorly of neutral words, it is considered as a suggestion.

b. Identification of incomplete information
   When the users don't provide the complete essential information to service their request (as in Example given in Figure 2), we call it as incomplete information. We restrict ourselves to identification of this type of complaints only. This can be done as a next step after identification of the type of tweet. If it is a complaint, the keywords or named entities extracted from the tweet text can be analyzed to find out whether the extracted information is sufficient as per the Railways Parameters. For example, if the complaint is about some ongoing train, the complaint should have a PNR number, rest all information about the customer can be taken from the existing software. If the complaint is about some failed transaction, then the complaint should have some value for user id, mobile number, transaction number, etc.

c. Identification of actionable tasks
In this part of the research, we aim to categorize the complaints into the categories as used by the existing software. Some example categories are Divyangjan Facilities, Bed Roll, Staff Behavior, Cleanliness, Passenger amenties, Coach-Maintenance, Water Availability, Unreserved Ticketing, Catering & Vending Service to mention a few.

d. Identification of the Division, Region and Department for Redressal
Once the complaint is categorised, it can be assigned to the relevant Division, Region and Department with minimum human intervention.

## 9. Novelty of the Project
We have found some of the related researches which work around similar situation. In [Kruspe et. al, 2021], authors have identified the importance of social media in crisis situations. They emphasize on automatic analysis of messages to extract crucial information. They present a review of detection problems based upon three research techniques *viz*. filtering using keyword and locations, crowdsourcing and machine learning. In [Preotiuc-Pietro et.al, 2019 ], authors have presented a linguistic analysis of complaining as a speech act. They have used neural models of complaints using distant supervision. The authors have focused on identification of syntactic patterns and linguistic markers specific of complaints.
Railway Complaint Tweets Identification has been taken up in [Akhtar & Beg, 2021]. The objective of this research paper is to classify tweets into complaint or suggestion.
Taking motivation from such researches, we have refined a novel objective. We focus on identification the nature of tweet(complaint, suggestion or appreciation). The novelty also lies in leveraging the power of text analytics for identification of actionable tasks if the tweet is a complaint. We also identify the incomplete information, i.e. whether the minimum required information has been provided in the tweet. If the incomplete information is there, prompt the customers to provide the complete information. We further identify the actionable tasks from the text analysis of tweet complaints, which can then be assigned to the relevant division, region and department.

## 10. Strategy
The practice currently being followed for responding to tweets addressed to Railways is manual analysis. It is apparent that adding automation to the task will help ease the human effort, improve response time, and may even allow detection of a crisis situation.
Text analytics is a combination of powerful techniques for automatic identification of crucial information from raw text. In this research project, raw text consists of the tweets posted on twitter handles of railways and its officials. The proposed approach will benefit both the railways(by improving the response time and decreasing the human effort) and the customers (by providing instant response).

## 11. Target Audience
1. Twitter Users
All those customers who use twitter for communication with Indian Railways. The twitter users can benefit in terms of first response because the software plugging will speed up the tweet processing.

2. Railways Twitter Cell
Since the software plug-in that we propose to build in this project aims at reducing the human effort involved in the analysis of the tweets. The persons working in the Twitter cell of Indian Railways will benefit in terms of reduction in volume of data that they currently handle.

## 12. Technology and Research Methodology

Natural Language Processing (NLP) applied in Python language makes the implementation most feasible and efficient. NLP shall understand and provide area of interest/impact of inputs whereas Python provides a set of instructions to the machine which construct such NLP logic.

Python language provides some efficient libraries for implementation of high level processing in spoken languages; this gives it an upper hand over other modern languages like JAVA/ C#.
Time complexity (better in C++) is also compensated by better code complexity. Moreover, code length is reduced to a considerable amount, decreasing effort in software development and its maintenance. Addition and future improvements in the software is much easier through this approach of least code complexity.

Research methodology in NLP is based on selection of predetermined keywords and their combinations (like 'good' and 'not good') present in the input and then classification on the basis of this selection (depending on frequency of occurrences) to categorize the text as 'complaint', 'suggestion' or ' appreciation' and further into concerned departmental interests.

## 13. Time Schedule of activities giving milestones

## 14. Suggested plan of action for utilization of research outcome expected from the project

## 15. Deliverables:

## 15 a. Publications, Technologies, Products, IPRs

## 15 b. Any other research output

## 15 c. Enterpreneurship Development/Techno-Commercial incubation

## 15 d. Start-ups and Spin off companies

## 15 e. Employment Generation

## 16. Cost Benefit Analysis

## 17. Budget:

| Item | | | Budget(lakhs) | | | Total(in Rupees) |
|---|---|---|---|---|---|---|
| A | **Recurring** | | | | | |
| | 1. Salaries/wages | | | | | |
| | Designation & Number of Persons | Monthly Emoluments | | | | |
| | | | | | | |
| | | | | | | |
| | 2. Consumables | | | | | |
| | 3. Travel | | | | | |
|   | 4. Contingency |   |   |   |   |   |
|   | 5. Other Costs: |   |   |   |   |   |
| B | Equipment(minor computational resources) |   |   |   |   |   |
|   |   |   |   |   |   |   |
| C | **Total(A+B)** |   |   |   |   |   |
| D | **Overhead(10%)** |   |   |   |   |   |
| E | **Grand Total(A+B+C+D)** |   |   |   |   |   |
| F | **Industry Support** |   |   |   |   |   |
| G | **Fund Requested from the TIH** |   |   |   |   |   |
|   |   |   |   |   |   |   |

**18. Expertise**
**18.1 Expertise available with the investigators in executing the project**
**18.2 Key publications of the Investigators pertaining to the theme of the proposal and TIH during the last 10 years**
**18.3 Infrastructural Facilities available with the investigator**
**18.4 Equipment available with the Institute/Group/Department/ for the project:**

**19. Endorsement letter from the PI and Co-Pis Institute**

**20. Duration**